\documentclass[preprint,12pt]{elsarticle}

\usepackage{xcolor}
\usepackage{wrapfig}
\usepackage{url}

\usepackage{amssymb}
\usepackage{amsmath}

\usepackage{lineno}

\newcounter{bla}

\journal{Computer Physics Communications}
\newcommand{\tadah}{\textit{Tadah!} }
\begin{document}

\begin{frontmatter}



\title{\tadah A Swiss Army Knife for Developing and Deployment of Machine Learning Interatomic Potentials}


\author[a]{Marcin Kirsz}
\author[a]{Ayobami Daramola}
\author[a]{Andreas Hermann}
\author[b]{Hongxiang Zong}
\author[a]{Graeme J. Ackland\corref{author}}

\cortext[author] {Corresponding author.\\\textit{E-mail address:} gja@ed.ac.uk}
\address[a]{Centre for Science at Extreme Conditions and SUPA, School of Physics and Astronomy, University of Edinburgh, Edinburgh EH9 3FD, United Kingdom}
\address[b]{State Key Laboratory for Mechanical Behavior of Materials, Xi’an Jiaotong University, Xi’an 710049, China}

\begin{abstract}
The \tadah code provides a versatile platform for developing and optimizing Machine Learning Interatomic Potentials (MLIPs). By integrating composite descriptors, it allows for a nuanced representation of system interactions, customized with unique cutoff functions and interaction distances. \tadah supports Bayesian Linear Regression (BLR) and Kernel Ridge Regression (KRR) to enhance model accuracy and uncertainty management. A key feature is its hyperparameter optimization cycle, iteratively refining model architecture to improve transferability. This approach incorporates performance constraints, aligning predictions with experimental and theoretical data. \tadah provides an interface for LAMMPS, enabling the deployment of MLIPs in molecular dynamics simulations. It is designed for broad accessibility, supporting parallel computations on desktop and HPC systems. \tadah leverages a modular C++ codebase, utilizing both compile-time and runtime polymorphism for flexibility and efficiency. 
Neural network support and predefined bonding schemes are potential future developments, and \tadah remains open to community-driven feature expansion. Comprehensive documentation and command-line tools further streamline the development and application of MLIPs.
\\


\noindent \textbf{PROGRAM SUMMARY}

\begin{small}
\noindent
{\em Program Title:} Tadah!                                         \\\\
{\em CPC Library link to program files:} (to be added by Technical Editor) \\\\
{\em Developer's repository link:} https://git.ecdf.ed.ac.uk/tadah \\\\
{\em Licensing provisions:} GPLv3 \\\\
{\em Programming language:} C++                                  \\\\
{\em Supplementary material:} Installation instructions and usage examples are available in the \tadah online documentation at https://tadah.readthedocs.io                               \\\\
{\em Nature of problem:}
Atomistic modelling, particularly molecular dynamics,  is among the most popular techniques used in physics and chemistry research.  Accurate and efficient methods are required to generate forces for such simulations.  Quantum mechanical calculation of the electronic structure is the ``gold standard" here, but is restricted to relatively small systems.  Hence, interatomic potentials have a role in allowing large scale simulations, provided they have adequate accuracy.

Over the past two decades, the paradigm has shifted from developing interatomic potentials using physics-informed functional forms to generating machine learning interatomic potentials (MLIPs) with more flexible mathematical forms, albeit lacking physical meaning.

MLIPs typically lack physical insights in trained models, requiring comprehensive datasets from methods like density functional theory during model parametrization. While training on energies and derivatives may ensure good interpolation within the training data, it frequently lacks transferability, failing environments unlike those in the training data. At the same time the ability to fit a model to experimental data increasingly seems to be lost.

In essence, the MLIP framework includes a feature vector (descriptor/fingerprint) and a regression method (e.g., kernel ridge regression or neural networks). Currently, there are  numerous frameworks, but there is no consensus on a superior approach. Competing factors such as accuracy, transferability, efficiency, and dataset requirements influence user decisions.  
Because of the fragmentation of the methodologies, users frequently need to compile multiple software tools, which then often prove to be not end-user-ready.  

A niche exists for a more flexible framework within which users can mix and match descriptor and regression methods.  This would allow MLIP developers to experiment, test, and rapidly deploy models to production, using the methods best suited to their system, application and desired outputs.

It is  desirable to provide additional tools for fitting and verification of models. Furthermore, deployment in molecular dynamics (MD) settings, requires effort to integrate the MLIP with third-party MD software. Plugins for standard MD codes should be available alongside MLIP fitting codes. 
\\

{\em Solution method:}
\tadah is a unified program designed for the development of advanced machine learning interatomic potentials using various descriptors, cutoffs, and regression methods. It also allows rapid deployment of these through the MD LAMMPS plugin [1]. It supports mono- and multi-species systems as well as advanced (custom) fitting procedures. The software provides unique pathways to embed prior domain knowledge into a model and offers a range of additional tools for dataset conversion, manipulation, and plotting.

\tadah allows users to incorporate their physical insight by combining multiple descriptors into a single composite descriptor. Each constituent descriptor can be further customized by specifying a unique cutoff function and distance. Additionally, users can define specific chemical species pairs that each constituent descriptor will target, enabling tailored interactions within the model.

\tadah also provides users with a nested fitting procedure that allows the optimization of model hyperparameters with a custom-defined loss function. In principle, the loss function can incorporate any quantity obtainable from MD simulation and be compared against experimental values for scoring.

The \tadah toolkit provides a command-line interface (CLI) binary as well as a C++ API for advanced use. The majority of features are available via the CLI, while the API allows developers and users to rapidly implement and test new descriptors and various regression strategies. It supports seamless integration of new types of descriptors, with an API that enables their addition. We follow the philosophy: ``Implement once, use it everywhere across \tadah and LAMMPS." The code is open source, and we aim to drive development based on user feedback. We encourage users to request new features and provide feedback. We are also open to individual contributions and collaborations.
  
{\em Additional comments including restrictions and unusual features:}
\tadah provides two end-user packages: {\it Tadah!MLIP} for the development of MLIPs and {\it Tadah!LAMMPS} for deployment in molecular dynamics settings. The {\it Tadah!MLIP} software can be built as a desktop version using OpenMP. For users working with large datasets or complex descriptors, a massively parallel version of {\it Tadah!MLIP} is available with MPI, designed for HPC architectures.
   \\
   
\textbf{Keywords}: Machine learning interatomic potentials, Molecular dynamics, Scientific machine learning, Physics aware machine learning, Computational modelling.
\\

\end{small}
   \end{abstract}
\end{frontmatter}
\section{Introduction}

Atomic-level simulations rely heavily on the choice of interatomic potentials. While empirical potentials derived from experimental data are computationally efficient, they often lack accuracy. In contrast, ab initio potentials, grounded in quantum mechanics, provide robust and transferable predictions, albeit at a greater computational cost. The emergence of machine learning interatomic potentials (MLIPs) allows for the design of models that combine computational efficiency with the capacity for high accuracy \cite{Deringer2019}.

Numerous software packages have been developed for training MLIPs \cite{Batatia2022mace,Bartok2010,Yanxon2021,DeePMD,DScribe,Lysogorskiy2021,Novikov2021,SchNetPack,TorchANI,Rupp2024}. Python-based tools support diverse descriptors but are inherently slower. Conversely, Fortran and C/C++ packages offer speed but are often restricted to a single descriptor type or regression method. Although Python codes can be sped up using external routines, the escalating complexity of fitting routines eventually limits these optimizations. The current landscape is dominated by packages employing neural networks as their preferred regression method.

Constructing an MLIP begins with prior knowledge. This involves selecting the functional form for the feature vector (descriptor), regression methods, training database, and machine learning (ML) hyperparameters (HP). These decisions must be made before initiating regression. These choices control the model's complexity and its capacity for accuracy, computational efficiency, and transferability between different atomic environments.
There is typically a manual and iterative process where the model is trained on the training data and fine-tuned with the validation set \cite{YANG2020}. Using different pieces of software is suboptimal from the user's perspective, as it significantly limits the ability to experiment. Moreover, optimizing hyperparameters is crucial during model development. Surprisingly, to the best of our knowledge, this topic remains inadequately addressed within the MLIP community \cite{Fiedler2022,duToit2023,duToit2024}.

In this work, we present novel, semi-automated training methods designed to facilitate the development of general-purpose MLIP models \cite{Kirsz_N2_2024,Pruteanu2021}. These methods are implemented in the \tadah software. The software is fully interfaced with LAMMPS \cite{LAMMPS}, allowing for efficient large-scale molecular dynamics simulations.

\section{Methods and Discussion}
The goal of interatomic potentials is to represent the potential energy surface (PES) for a given set of atomic coordinates. MLIPs extend the classical approach by decomposing the system's total potential energy, $U_{tot}$ into local atomic energies  $U_i$ for each atom. 

\begin{figure}[ht]
\centering
    \includegraphics[width=0.3\textwidth]{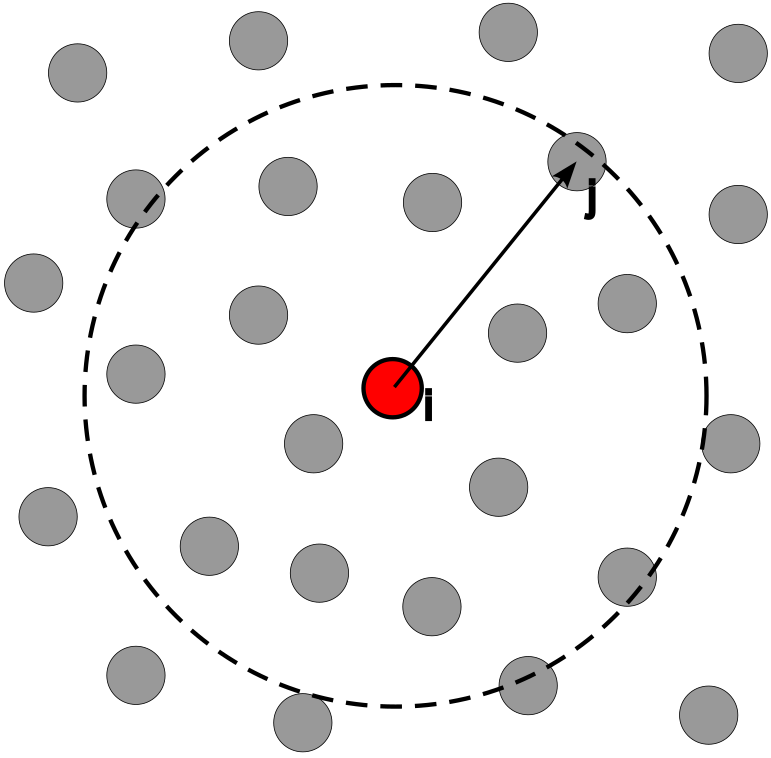}
    \caption{Local atomic environment of atom $i$. The atom $j$ is considered as within the local environment of atom $i$ if the separation between atoms is within user defined cutoff distance $r_c$.
    \label{fig:local_atomic_env}}
\end{figure}

\begin{equation}
    U_{tot}=\sum_i^N U_i = \sum_i^N \mathcal{M}(\mathbf{d}_i)
    \label{eq:loc_at_env}
\end{equation}

In eq. \ref{eq:loc_at_env}, a descriptor vector, $\mathbf{d}_i$, captures information about the local atomic environment of the $i$th atom  within a predefined cutoff radius $r_c$. When integrated into a trained ML model, $\mathcal{M}$, it predicts the local energy $U_i$. The summation is over all atoms in the simulation. The forces can be obtained by taking the derivatives of $U_{tot}$ with respect to the atomic coordinates.

The \tadah software supports multiple types of descriptors and regression methods, as well as the ability to construct composite descriptors. This section provides a brief overview of the theory behind MLIPs. The features unique to \tadah are discussed in more detail. In particular, we present the generalized form of the descriptor vectors currently supported by \tadah and the semi-automated nested fitting procedure designed to reduce manual effort during model parametrization and improve transferability.

\subsection{Parameters} 
\label{SS:PM}
Both $\mathcal{M}$ and $(\mathbf{d}_i)$ are functions\footnote{strictly, $\mathcal{M}$ is a functional and $(\mathbf{d}_i)$ is a vector of functions}.  They contain two types of parameters, {\it Learned parameters} (LPs) which are optimised in the training process (e.g. regression coefficients) and {\it hyperparameters} (HPs) which are not (e.g. cutoffs, and even the length of the descriptor vector).  \tadah offers an external loop which can optimise the hyperparameters either to improve the fitting, or to optimise some emergent property such as equation of state.

\subsection{Generalised Descriptor} 
\label{SS:GD}
The descriptor vector ($\mathbf{d}$) captures information about the local atomic environment of an atom. To construct $\mathbf{d}$, we first define a set of hyperparameters that specify its functional form. Note that vector component $\mathbf{d}^{(p)}$ of $\mathbf{d}$ may have the same functional form with a different set of HPs. 

For example, one can construct a simple Lennard-Jones type descriptor with two vector components, corresponding to attraction and repulsion respectively, and consider the exponents of particle separation $r$ as the following sets of hyperparameters: $\{-6\}_0$ and $\{-12\}_1$. Additional examples may include descriptors that use Gaussian functions, where hyperparameters might define the width and position of the Gaussians. Chemical species-dependent weights can also be used to tailor the descriptor for specific interactions, enhancing its sensitivity to the chemical composition of the environment.

The \tadah software supports two- and many-body types of descriptors. The specific angular type descriptor is omitted, as three-body interactions can generally be incorporated into a many-body format, This approach avoids the cost associated with the computation of angular descriptors \cite{Zhang2019}. The list of currently supported types of descriptors is available in online documentation \cite{tadah_docs}.

\subsubsection{Two-body components in the descriptor}

The simplest form of component in the descriptor is a two-body
functional form.  It is motivated by the idea that the energy depends on pairwise interactions between atoms.
A two-body $p$-th component of the descriptor of the $i$th atom is
\begin{equation} \label{eq:2b_desc}
    \mathbf{d}^{(p)}_i = \sum_{j\neq i} B^{\{\zeta\}_p}(r_{ij}) f_c^{\{\zeta\}_p}(r_{ij})
\end{equation}
where $B$ is a descriptor specific function with a set of hyperparameters ${\{\zeta\}_p}$, $f_c$ is a cutoff function which ensures that energy goes smoothly to zero at the cutoff distance, and $r_{ij}$ is a separation between $i$ and $j$ atoms. The summation is over all neighbours of central atom $i$ which are within $r_c$. Note that $r_c$ is included in the $\{\zeta\}_p$ subset of hyper parameters as it might be required by the $B$ functions as well.

\subsubsection{Many-body  components of the descriptor}
The many-body descriptors of the $i$th atom are motivated by the idea that the energy depends on the local density. 
 They satisfy the following form 
\begin{equation} \label{eq:mb_desc}
    \mathbf{d}^{(p)}_i = \mathcal{D}^{\{\zeta\}_p}\Big(\boldsymbol{\rho}^{\{\zeta\}_p}_i\Big)
\end{equation}
where the vector of atomic density $\boldsymbol{\rho}_i$ of the $i$th atom is built by expanding the density using the basis set of choice.
\begin{equation}
\label{eq:mb_density}
    \boldsymbol{\rho}_i^{\{\zeta\}_p} = \Big(\sum_j \psi_1^{\{\zeta\}_p}(\mathbf{r}_{ij}),  \dotsc, \sum_j \psi_{max}^{\{\zeta\}_p}(\mathbf{r}_{ij})\Big)
\end{equation}

For linear regression, the functional $\mathcal{D}$ in eq. \ref{eq:mb_desc} must ensure invariance under permutations of atoms of the same species, as well as inversion, translation, and rotation of the system. In principle, these invariances can also be incorporated into a custom non-linear model.

\subsubsection{Composite descriptors}
\begin{figure}[ht]
\centering
    \includegraphics[width=1.0\textwidth]{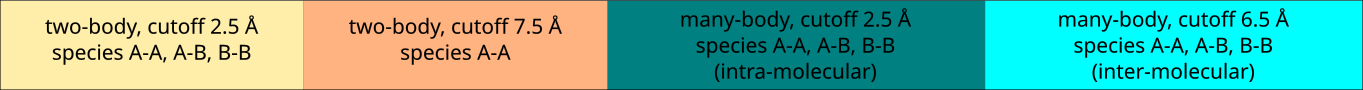}
    \caption{The schematic representation of the composite descriptor which allows the user to mix and match any combination of descriptors.
    \label{fig:composite_descriptor}}
\end{figure}
\tadah offers the flexibility to incorporate physical insights into the modelling process by choosing the physical functional forms of the descriptor components.
These multiple components are combined into a single, unified custom descriptor vector, referred to as a composite descriptor (fig. \ref{fig:composite_descriptor}). This approach allows for a more nuanced representation of the system's interactions. Each constituent descriptor within this composite can be further tailored by specifying a unique cutoff function and an interaction distance, enhancing the precision and relevance of the model. Furthermore, it is possible to define particular chemical species pairs that each constituent descriptor will target. This capability enables the creation of highly specific interactions, ensuring the model accurately reflects the complex dynamics of the system being studied. By leveraging these features, models can be developed that are not only highly customized but also deeply informed by domain-specific knowledge.

\subsection{Regression}

\tadah uses the regularized form of the normal equation during the regression stage to find the optimal parameters that minimize the error between predicted and actual data:

\begin{equation} \label{eq:w}
\mathbf{w} = (\mathbf{X}^T \mathbf{X} + \lambda \mathbf{I})^{-1} \mathbf{X}^T \mathbf{t}
\end{equation}

where \(\mathbf{w}\) is the optimized weight vector, \(\mathbf{X}\) is the design matrix which is constructed using descriptor vectors, \(\mathbf{t}\) is the target vector consisting of energies, forces, and stresses, \(\lambda\) is the regularization parameter, and \(\mathbf{I}\) is the identity matrix. This closed-form solution is equivalent to minimizing the loss function, defined as the sum of squared errors, with the advantage of being exact.

An evidence approximation algorithm is used to estimate $\lambda$, which helps prevent model overfitting by shrinking the weights. Users also have the option to select this term manually. Models regularized with this approach may exhibit higher bias on training data but achieve lower variance and better accuracy on test data.

\tadah currently supports two regression methods: Bayesian Linear Regression (BLR) and Kernel Ridge Regression (KRR) \cite{Bishop2006}.
The key difference between KRR and BLR lies in the computation of the design matrix \(\mathbf{X}\). Once constructed, both methods apply regularized linear regression (eq. \ref{eq:w}).
For KRR, the design matrix \(\mathbf{X}\) is replaced by the sparse kernel matrix \(\mathbf{K}\), constructed using a kernel function \(k(\mathbf{x}_i, \mathbf{x}_j)\). Here, the index \(i\) goes over all descriptors in the training dataset, while index \(j\) goes over all preselected basis vectors.
For BLR, each row of \(\mathbf{X}\) is post-processed with the basis function of choice, resulting in the \(\Phi\) matrix.
For both algorithms, the design matrix can include descriptor vectors for structural potential energies and optionally for forces and stresses.
Various kernel types and basis functions are supported. For details, refer to the online documentation \cite{tadah_docs}. 

In BLR, regression is framed as a probabilistic model, allowing for robust predictions that incorporate uncertainty. BLR assumes a prior distribution over the weight vector \(\mathbf{w}\). The model updates this prior to a posterior distribution based on the observed data. This approach not only predicts outputs but also provides a measure of uncertainty for predictions, which is valuable for assessing model confidence.

Our KRR implementation leverages the Empirical Kernel Map (EKM) \cite{EKM}, enabling the processing of large datasets and producing sparse outputs. EKM kernelizes algorithms that use vectors by projecting new vectors into a feature space defined by basis vectors and a kernel function. This allows standard (not kernalized) algorithms to operate without modifications. \tadah provides several tools for selecting a suitable set of basis descriptor vectors, such as simple random selection or recursive finding of linearly independent vectors in a kernel space \cite{LISF,dlib}. Thanks to EKM, tools from BLR can be applied to the sparse matrix \(\mathbf{K}\), allowing for the prediction of errors and uncertainty estimation in the model.

\subsection{\label{hpo} Nested Fitting Procedure}

One of the major challenges for MLIPs is their poor transferability beyond the training dataset. \tadah addresses this by incorporating a hyperparameter optimization cycle. Unlike standard approaches that primarily fit forces and energies, \tadah employs an iterative procedure to generate ``trial" potentials with varying hyperparameters. These are applied via LAMMPS to evaluate macroscopic properties, fitting them to theoretical or experimental data. The best-performing\footnote{Either in terms of the fit to training data, execution speed, or a custom user-defined metric.} combination of descriptors and kernels are chosen resulting in increased performance, transferability, and applicability. 

In developing MLIPs, both LPs and HPs are crucial. The model architecture, including the choice of ML algorithm and descriptors, reflects prior knowledge and is fine-tuned through HP optimization. The global objective function, known as the evaluation function in ML literature, plays a central role.

The aim of HP optimization is to minimize the global loss function \(\mathcal{L}(\mathcal{M}, \mathcal{T})\), which is a common choice for a global objective function, where \(\mathcal{M}\) is the model and \(\mathcal{T}\) is the training dataset.

\begin{equation}
    \theta^{*} = \underset{\theta \in \Theta}{\mathrm{arg} \, \mathrm{min}}  \, 
    \mathcal{L(M,T)}
\end{equation}

Here, the model is parameterized by a set of HPs $\theta$, and the goal is to find the $\theta^{*}$ which minimize  $\mathcal{L(M,T)}$  within a search space $\Theta$.

HP optimization differs from other optimization problems. While it is theoretically possible to obtain the gradient of the loss function with respect to the HPs, in practice, this is rare due to discontinuities or non-differentiable search surfaces. Care must be taken to avoid overfitting HPs to a particular dataset, as different datasets may require different optimal HP values.

In general, MLIPs achieve high accuracy for constructing PES for local atomic configurations similar to training data, owing to using generic functions with numerous free parameters. Traditional MLIP model development involves adjusting HPs during manual iterative training and validation, followed by final testing. This approach is simplistic, resource-intensive and time-consuming.

The \tadah global loss function (GLF) is defined as:

\begin{equation}
\label{eq::HPO::methods::GLF}
    \mathcal{L}_g(\mathcal{M},\mathcal{T}) = \sum_\alpha \omega_\alpha \mathcal{L}_\alpha(\mathcal{M},\mathcal{T})
\end{equation}

In this equation, \(\mathcal{L}_\alpha\) represents the model error associated with the \(\alpha^{th}\) constraint, and the \(\omega_\alpha\) are weighting parameters that indicate the importance of $\alpha$ in the fitting procedure. The constraint loss function takes the form:

\begin{equation}
    \mathcal{L}_\alpha(\mathcal{M},\mathcal{T}) = \big|\mathcal{P}_\alpha(\mathcal{M},\mathcal{T}) - t_\alpha \big|^N
\end{equation}

Here, \(\mathcal{P}_\alpha\) denotes the prediction for the \(\alpha^{th}\) constraint, with \(t_\alpha\) as the target value. The power \(N\) determines the type of loss function: setting \(N=1\) results in an absolute loss, while \(N=2\) yields the commonly used quadratic loss function.

The weight factor \(\omega_\alpha\) has  units of the corresponding constraint \(\alpha\) raised to the \(N^{th}\) power. This ensures that the product in eq. \ref{eq::HPO::methods::GLF} is unit-less. The interpretation of the weighting parameters is intuitive—they control the numerical precision of the obtained loss for a given \(\alpha\) relative to other weights. For example, doubling a weight makes its associated constraint twice as important as before.

The GLF evaluates the model's performance not just on the validation set but also by incorporating performance constraints (PC) on the model's physical predictions. These constraints enhance the predictive power of the interatomic potential. PCs can be divided into two types: the RMSE fit to target vector {\bf t} (eq. \ref{eq:w}), and those that are physically motivated, such as the model's ability to reproduce specific surface energies or energy differences between crystal structures. \tadah provides a flexible interface that allows users to use custom LAMMPS scripts to evaluate the PCs of their choice. 

In contrast, search space constraints (SSC) define the configurational space for HPs, which the optimization algorithm explores to satisfy the PCs. 
SSCs are applied more directly to the model's architecture, influencing parameters like the positions and widths of Gaussians in a descriptor or a cutoff distance. 

\begin{figure}[ht]
\centering
    \includegraphics[width=1.0\textwidth]{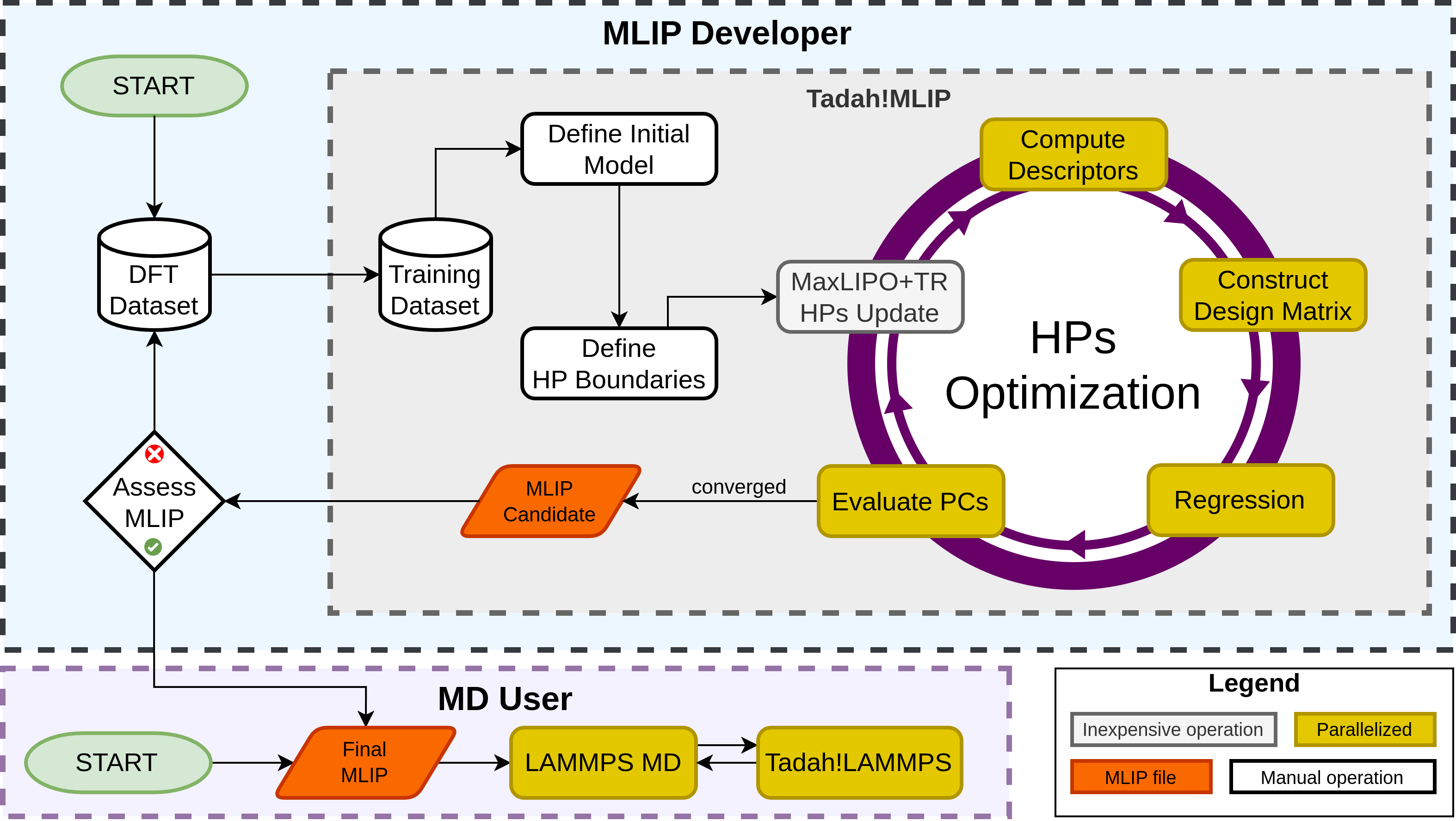}
    \caption{The hyperparameter optimization routine as implemented in {\it Tadah!}. The MLIP development starts with constructing a DFT dataset. Users can employ existing datasets or build new ones using external tools. The training stage involves sampling from the DFT data to create a training dataset with the {\it Tadah!MLIP} toolkit. Next, the MLIP developer defines the model and selects which HPs to optimize, along with setting appropriate search space and performance constraints. Once configured, the automated optimization process begins, generating ``trial" potentials and evaluating them against PCs. Upon completion, the MLIP candidate can be manually evaluated by the developer before being distributed to MD users. {\it Tadah!LAMMPS} is an independent plugin of {\it Tadah!MLIP} and is required to run MD simulations with the final MLIP.}

    \label{fig:hpo_cycle}
\end{figure}

The global optimization algorithm (GOA), as illustrated in fig. \ref{fig:hpo_cycle}, works iteratively to optimize model architecture with respect to training data, validation sets, search space, and performance constraints. The training and validation data sets are constructed by the user and remain unchanged throughout the process, as \tadah currently lacks tools for automating this step. The success of the potential depends on the quality of the training data and \tadah allows the user to tune the training data to their intended application. The primary goal of the GOA is to enhance model architecture and, consequently, its transferability.

Optimization begins by defining target PCs and SSCs in a configuration file,  where the user assigns a weight to each PC indicating its importance. The automated iterative process involves: selecting candidate HPs from SSCs, training the model with new settings, and evaluating performance against PCs. This cycle repeats until convergence criteria, like a specific GLF value, are met, or until manually stopped. The potential is refined through both changes in HPs and values of $\mathbf{w}$.

The HP selection process is managed by the \textit{MaxLIPO+TR} algorithm from the Dlib C++ library \cite{dlib}. An enhancement of the original LIPO algorithm \cite{Malherbe2017GlobalFunctions}, it estimates the Lipschitz constant to construct an upper bound to the objective function, optimizing towards a global maximum. It evaluates points randomly, comparing their upper bounds to find improvements. To address slow convergence near optima, it employs a trust region method (+TR), assuming a quadratic surface to swiftly converge on local optima \cite{Goldfeld1966MaximizationHill-Climbing, Sorensen1982NewtonsModification}.

\section{Implementation}

\begin{figure}[ht]
\centering
    \includegraphics[width=0.5\textwidth]{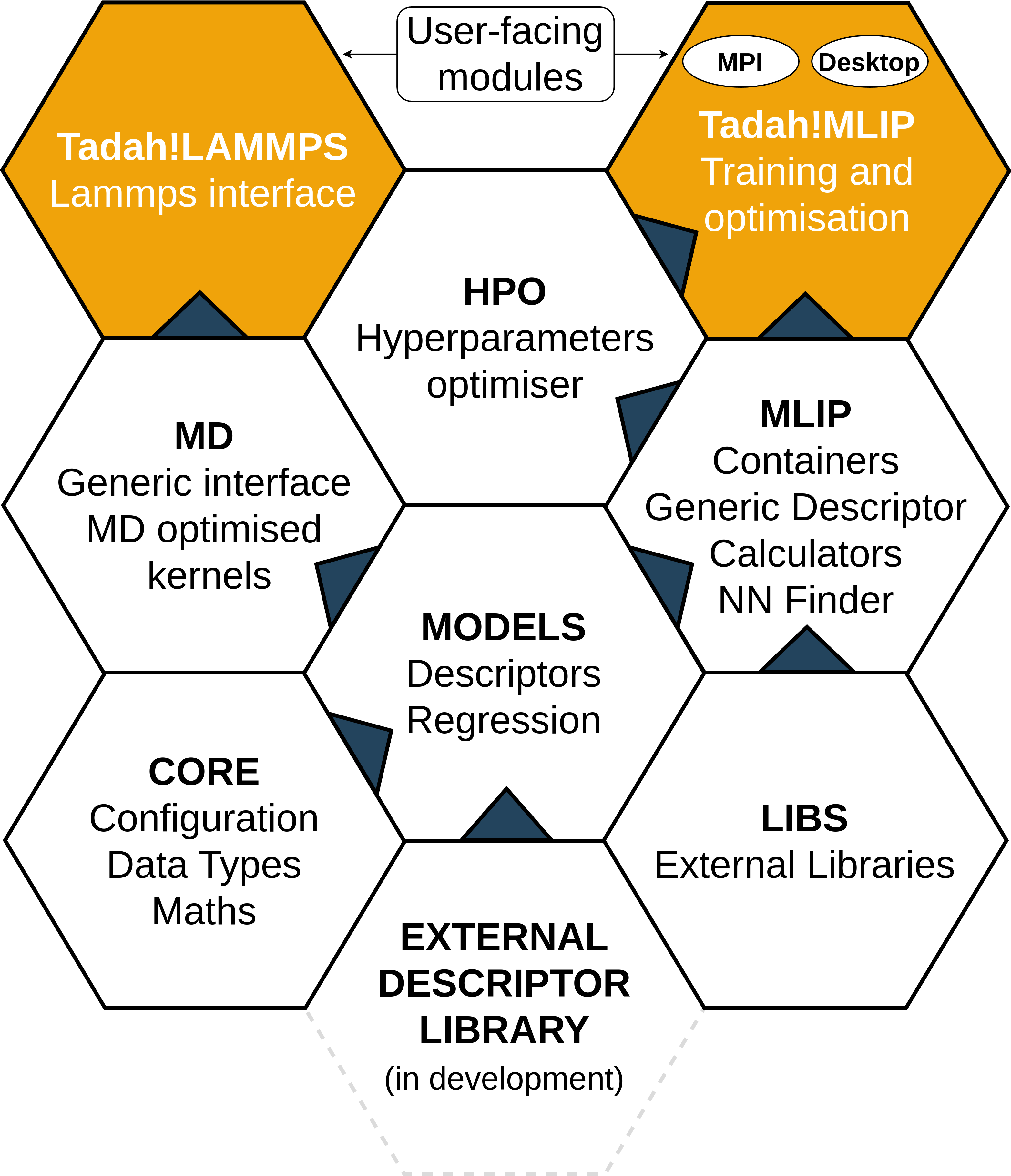}
    \caption{The \tadah codebase is written in object-oriented C++ and consists of six independent modules, combined to create two user-facing modules. {\it Tadah!MLIP} is designed for training and optimizing MLIPs, while {\it Tadah!LAMMPS} deploys them in the MD setting. This modular approach enhances performance, code reusability, and reduces maintenance.}
    \label{fig:tadah_modules}
\end{figure}

The \tadah code is open-source and implemented in C++. Developed using Git, it promotes collaboration through a modular structure (see fig. \ref{fig:tadah_modules}). The software consists of six independent modules that support the user-facing components, {\it Tadah!MLIP} and {\it Tadah!LAMMPS}, enabling efficient code reuse. To enable flexible usage as a stand-alone library, it employs generic programming techniques, like class templates, providing compile-time polymorphism for efficient and reusable components.

For command-line interfaces and the LAMMPS plugin, which require run-time polymorphism, \tadah uses a factory method design pattern. This approach allows for selecting model components based on configuration files, combining compile-time efficiency with run-time flexibility. 
The object-oriented design focuses on performance, using shallow abstraction layers for critical functionalities like descriptor computation to maintain efficiency. API documentation is generated with Doxygen and published online on ReadtheDocs for easy access \cite{tadah_docs}.

Our software development focussed on robustness; unit, functional and integration tests are supported by GitLab continuous integration and continuous delivery (CI/CD) to automate processes, catch bugs early, and ensure compatibility with new LAMMPS versions. This approach enables rapid, reliable software releases and simplifies collaboration and the addition of new features.

{\it Tadah!MLIP} can be compiled on desktops and high-performance computing (HPC) facilities. The desktop version is parallelized with OpenMP and can handle datasets of tens of thousands of configurations. For more demanding tasks, the MPI version, specifically designed for HPC architectures, fully parallelizes descriptor computation and regression using an MPI host-client design pattern, managing extremely large datasets necessary for accurate MLIP parameterization.

\subsection{Dependencies and availability}
The software is available under the GPLv3 license from \url{https://git.ecdf.ed.ac.uk/tadah}, and the online documentation is hosted at \url{https://tadah.readthedocs.io}. The LAMMPS interface and {\it Tadah!MLIP} toolkit require C++11 and C++17 compatible compilers, respectively. The code has been successfully deployed on a range of Linux architectures, from desktop versions like Alpine or Ubuntu to the HPE Cray Linux Environment, as well as macOS. Fortran routines from LAPACK \cite{lapack} and ScaLAPACK \cite{scalapack} (for the \tadah MPI version only) must be available on the user's system. \tadah utilizes CMake for configuration and building of its components. Git and an internet connection are required during the installation process.

The code has not been tested on Windows; however, we expect the LAMMPS plugin to be compatible. For Windows users, we recommend compiling either {\it Tadah!MLIP} or {\it Tadah!LAMMPS} using a Linux virtual machine. 

The software employs several popular libraries such as Dlib \cite{dlib}, Boost, toml11, and CLI11, which are either contained within the \tadah codebase or automatically downloaded and configured during the installation process when needed for a build.

\section{Typical Usage}

\subsection{Using \tadah Potentials}

Many users are primarily interested in utilizing pre-trained potentials rather than developing them. For this purpose, the {\it Tadah!LAMMPS} plugin is all they need, as it allows these potentials to be seamlessly integrated with LAMMPS like any other interatomic potential. The potentials are distributed as ASCII files and are generated by the {\it Tadah!MLIP} software.

To use this feature, users must \texttt{git clone} the LAMMPS plugin from the \tadah repository into the \texttt{lammps/lib} directory. The remaining compilation process is straightforward and follows the standard LAMMPS library and package installation procedures. For detailed steps, refer to the \tadah documentation \cite{tadah_docs}.
Once compiled, users can invoke \tadah potentials using the standard LAMMPS syntax:

\begin{verbatim}
pair_style      tadah
pair_coeff      * * pot.tadah ELEMENT1 ELEMENT2
\end{verbatim}
where, \texttt{pot.tadah} is the filename of the interatomic potential.

\subsection{MLIPs Development Toolkit}

This section provides a brief overview of the {\it Tadah!MLIP} toolkit, designed specifically for the development of Machine Learning Interatomic Potentials (MLIPs).
The primary entry point to \tadah is the command-line program \texttt{tadah}, which offers various subcommands. These subcommands may require simple configuration files to facilitate MLIPs development. For examples and detailed explanations, please refer to the online documentation \cite{tadah_docs}.

Key subcommands include:

- \texttt{db}: Offers dataset manipulation functionalities such as joining, splitting, finding duplicate structures, and selecting subsets.
- \texttt{train}: Performs regression training on a given dataset.
- \texttt{predict}: Predicts energy, forces, and stresses.
- \texttt{hpo}: Conducts hyperparameter optimization and training using nested fitting procedure as described in \ref{hpo}.
- \texttt{swriter}: Dumps dataset configurations into formats like CASTEP \texttt{.cell}, VASP \texttt{POSCAR}, or LAMMPS data files.
- \texttt{desc}: Calculates descriptors and saves them to a file.
- \texttt{convert}: Assists in extracting relevant data (atomic positions, energies and so on) from CASTEP (\texttt{.md}, \texttt{.geom}, or \texttt{.castep}) or VASP (\texttt{OUTCAR} or \texttt{vasprun.xml}) DFT calculations to construct training datasets.
- \texttt{plot}: Provides simple utilities to plot and visualize cutoff functions, basis functions (such as Gaussians), and the interaction energy between two atoms using a trained MLIP.

The CLI tools are well-documented and offer helpful descriptions when used with the \texttt{-h} or \texttt{--help} flags. For instance, \texttt{tadah db -h} provides guidance on dataset subcommands. Further documentation is available online.

\section{Limitations}
\tadah does not support predefined bonds; hence, everything is treated as atomic and intramolecular bonding must be learned from the dataset.  This means that bond breaking should be treated with extreme caution unless included in the training data.  \tadah also lacks schemes for incorporating  long-distance interactions. 

\subsection{Ongoing development}

\tadah remains under development.  The framework was designed to make it easy to implement enhancements.  

Notably, neural networks are currently not supported but could be plugged in to replace the regression step.
Implementing new descriptors such as 3-body, MACE or ACE that deviate from the functional forms of eq. \ref{eq:2b_desc} or eq. \ref{eq:mb_desc} will require modifications to the \tadah codebase. 

\section{Funding}
This work was supported by the UK national high-performance computing service, ARCHER2, for which access was obtained via the UKCP consortium and funded by EPSRC grant ref EP/X035891/1. M.K. was supported by the ARCHER2 eCSE software development programme, project ARCHER2-eCSE11-11.

\section{Rights Retention Statement }
For the purpose of open access, the author has applied a Creative Commons Attribution (CC BY) licence to any Author Accepted Manuscript version arising from this submission.

\section{Acknowledgments}
We would like to thank C.G. Pruteanu for their insightful reading and comments, and A. J. Iwasaki for testing the software and providing valuable feedback.

\section{Declaration of generative AI and AI-assisted technologies in the writing process}

During the preparation of this work the authors used Edinburgh University’s own platform ELM (Edinburgh Language Model) in order to improve the readability and language of the manuscript. After using this tool, the authors reviewed and edited the content as needed and take full responsibility for the content of the published article.





\bibliographystyle{elsarticle-num}
\bibliography{refs}







\end{document}